# Submicrometer Pattern Fabrication by Intensification of Instability in Ultrathin Polymer Films under a Water-Solvent Mix

*Ankur Verma and Ashutosh Sharma**

Department of Chemical Engineering and DST Unit on Nanosciences, Indian Institute of Technology Kanpur, Kanpur 208016, India

* Corresponding author: phone + 91-512-259 7026; Fax + 91-512-259 0104; e-mail ashutos@iitk.ac.in

ABSTRACT: Dewetting of ultrathin (< 100 nm) polymer films, by heating above the glass transition, produces droplets of sizes of the order of microns and mean separations between droplets of the order of tens of microns. These relatively large length scales are because of the weak destabilizing van der Waals forces and the high surface energy penalty required for deformations on small scales. We show a simple, one-step versatile method to fabricate sub-micron (>~100 nm) droplets and their ordered arrays by room temperature dewetting of ultrathin polystyrene (PS) films by minimizing these limitations. This is achieved by controlled room temperature dewetting under an optimal mixture of water, acetone and methyl-ethyl ketone (MEK). Diffusion of organic solvents in the film greatly reduces its glass transition temperature and the interfacial tension, but enhances the destabilizing field by introduction of electrostatic force. The latter is reflected in a change in the exponent, *n* of the instability length scale, $\lambda \sim h^n$, where *h* is the film thickness and *n* = 1.51±0.06 in the case of water-solvent mix, as opposed to its value of 2.19±0.07 for dewetting in air. The net outcome is more than one order of magnitude reduction in the droplet size as well as their mean separation and also a much faster dynamics of dewetting. We also demonstrate the use of this technique for controlled dewetting on topographically patterned substrates with submicrometer features where dewetting in air is either arrested, incomplete or unable to produce ordered patterns.

**Introduction**

There are extensive studies on the dynamics of instabilities and structure formation in thin polymer films because of their increasing applications in the areas of coatings (thermal and electrical insulation, optical, protective, etc.), plastic electronics including organic light emitting diodes (OLEDs), lubricants, adhesives, membranes and biological coatings etc.[1–6] In many such applications instabilities are undesirable, and various strategies have been proposed to enhance the stability of thin-films such as adding filler (nano)particles and other stabilizing additives.[7–11] Dewetting of ultrathin films (<100 nm) can give insights into the inter-surface forces that cannot be directly measured otherwise.[12–25,49,50] It also provides a handy toolbox to generate useful meso- and nano-scale patterns over large areas (~cm$^2$).[12–50] Spontaneous dewetting of ultrathin polymer films has thus been a subject of some excellent reviews owing to its scientific and technological importance.[23–25]

Dewetting on a flat homogeneous surface starts with the nucleation of randomly placed holes at a certain mean separation ($\lambda$), which grow and coalesce with time and result in randomly placed droplets of the polymer. Average diameter as well as the mean separation of dewetted structures are a function of initial thickness ($h$) and the interfacial tension of the polymer film.[12–25] However, randomness of the dewetted structure limits the usefulness of this method in some applications as a potential soft patterning technique. Various strategies have been explored to impose a long range-order in the dewetted structures. One strategy for the alignment of the dewetted structure is to combine it with other top-down lithographic approaches such as controlled dewetting on topographically or chemically patterned substrates. Dewetting of ultrathin polymers films on physico-chemically patterned substrates has also been extensively studied both theoretically[27–31] and experimentally.[32–48] However, despite its promising scientific and technological potential, the feature-size generated by the self-organized dewetting has two major limitations on the pattern resolution and its aspect ratio. The first limitation arises owing to the weak van der Waals destabilizing force and high surface tension ($\gamma$), both of which impose a severe limit on minimum feature size, which is related to the wavelength, λ of the long-wave instability in spinodal dewetting:[12–25]

$$\lambda = [-8\pi^2\gamma/(\partial\phi/\partial h)]^{1/2} \tag{1}$$

Where $h$ is the film thickness and $\phi$ is the destabilizing intermolecular potential ($\sim h^{-3}$ for van der Waals interaction).[12–25] In a model system of polystyrene (PS) thin-film on silicon substrate, this limits the spot size to >1μm even in case of films as thin as 10 nm. Other limitation is the very small contact angle (< 10°) and thus the aspect ratio of dewetted structures in air is rather small.

In a short communication,[48] we recently proposed a novel empirical method by which limitations on the feature size and aspect ratio can be overcome to a great extent. The PS thin (< 100 nm) films were destabilized by immersing into an optimal mixture of solvent (methyl-ethyl ketone and acetone) and water at room temperature. Selective diffusion of solvent molecules into the polymer matrix brings down its glass transition temperature ($T_g$) below room temperature and thus causes the instability and dewetting of the film. However, water being the majority phase in the bounding media inhibits the dissolution of PS. Room temperature dewetting in the liquid media provides the advantage of cleaner environment, greater control over the shape of structure and better defect control as compared to thermal annealing or dewetting in air. Further it prevents the sensitive materials from harsh conditions (UV, laser, electron beam, heating high vacuum etc.) of patterning. Interestingly, dewetting under the water-solvent mix reduces the droplet diameter by more than one order of magnitude compared to dewetting in air.[48]

The objectives of this work are: (1) to investigate the underlying mechanisms of intensification of surface instability under water-solvent mix including the length scale of instability before fragmentation of the film into droplets and, (2) to compare the conditions for the formation of ordered patterns on topographically structured substrates in air and under the water-organic mix. Toward these ends, we first systematically investigate the length scale of instability, droplet diameters and inter-droplet spacing as a function of film thickness (7–90 nm) in order to clearly differentiate these from the well-known aspects of spontaneous dewetting in air.[12–25] Finally, we explore the conditions for the formation of ordered patterns on topographically structured 1-D and 2-D substrates where dewetting occurs in highly confined but structured spaces. We thus demonstrate the use this technique for fabrication of ordered arrays of sub-micron polymeric lenses of tunable curvature by controlled dewetting under the water-organic mix. In particular, it is also shown that dewetting in air for the same films is unable to generate dewetting and ordered structures owing to the length scale limitations in confined spaces.

**Materials and Methods**

**Materials:** PS of average molecular weight ($M_w$) 280 kg mol$^{-1}$ and polydispersity index (PDI = $M_w/M_n$) of < 1.1 (Sigma Aldrich) in the form of pellets was used to make polymer solutions in HPLC grade toluene (Qualigens fine chemicals, India). Thin-films of thickness ranging from 7 nm to 100 nm were spin coated (Laurell Technologies, USA) at 3000 rpm for 1 minute on thoroughly cleaned Silicon (100) substrates (p-type, Waferworld, USA), with a native oxide layer of ~2 nm, by using 0.1–1.8 w/v % polymer solution in toluene. The spin coated films were left on the clean bench (ambient drying) for 2 hours and annealed in a mild vacuum oven at 60°C for 12 hours for the removal of solvent and to minimize the residual stresses. 1-D array of parallel open channels and 2-D array of circular/square pillars of 100 nm height and varying size and pitch were fabricated on silicon by the e-beam lithography (Xenos XeDRAW). PS films were coated on the EBL fabricated patterns directly using the method described above.

**Dewetting Methodology:** Annealed thin-films of PS were put in a dewetting chamber containing a liquid mixture of water, methyl-ethyl ketone (MEK) and acetone in the ratio 15:7:3. MEK is a good solvent for PS and acetone facilitates the mixing of water and MEK which are otherwise sparingly miscible and thus allows a clear single phase mixture. Diffusion of MEK in PS reduces its glass transition temperature and allows PS film to dewet at room temperature. We could confirm that this mixture has no significant solubility for PS as evidenced by negligible weight loss over 24 hours, which is much longer than the time for dewetting. Dewetting in air was initiated by placing thin-films in a chamber saturated with the same solvent vapors (MEK and acetone, 7:3). There was no significant difference if films were dewet instead by thermal annealing above its glass transition temperature. To examine the extent of dewetting under the water-solvent mix, the samples were taken out at different times and dried with hot air to freeze the structure. All the experiments were carried out in class 1000 clean room to minimize the chances of defect induced dewetting.

**Characterization:** Thickness of the spin coated films was measured by a nulling ellipsometer (Nanofilm, EP$^3$-SE) using a 532 nm green laser at an incident angle of 55° (Brewster's angle ~ 57°). Imaging of the dewetted structures was done using an optical microscope (Zeiss Axio observer Z1) in the bright field and also by a Field Effect Scanning Electron Microscope (FESEM) (Zeiss Supra 40VP) employing secondary electrons. The contact angle could also be imaged by the transverse view of the droplets in FESEM. Interfacial tension of PS in water-solvent mix and solvent vapor rich air is measured by pendant drop method using contact angle goniometer (ramé-hart, 190 CA).

## Results and Discussion

### Dewetting on flat substrates

Dewetting of PS thin-films on flat and smooth silicon substrate was studied in the water-solvent mixture and comparisons were made with the corresponding dewetting of identical thin-films in air saturated with solvent vapor. The evolution of dewetting was qualitatively similar in both the cases and similar to the dewetting behavior known previously in air.[12,13,15,19,20,46] Figure 1 summarizes the sequential dewetting stages in a 18 nm (Figure 1a–f) and a 60 nm (Figure 1g–l) thick PS films in the water-solvent mixture. In both the cases, onset of dewetting was observed in less than 5 seconds. Dewetting proceeded with the formation of isolated holes which grow in time and coalesce to form a network of polymer ribbons that eventually break to form isolated droplets as discussed first by Reiter for dewetting in air.[12,13] The sequence of morphological evolution was same as in air,[12,13,15,46] and similar to dewetting of PDMS films under water.[52] For the 18 nm thick PS film, dewetting was completed in less than 60 seconds, while for 60 nm thick film, complete dewetting took about 30 minutes. It may be noted that the time scales for dewetting under water-solvent mix are significantly shorter compared to the dewetting in air (~10 minutes for 18 nm and ~6 hours for 60 nm film). A much faster dynamics already points to intensification of instability, i.e., either by enhancing of destabilizing force or by reduction in the stabilizing force or both.

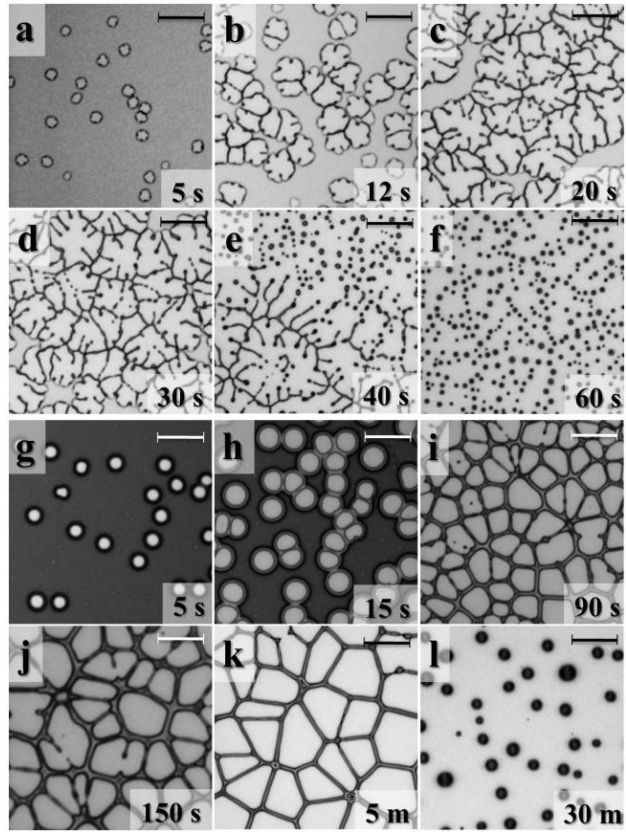

**Figure 1.** Time evolution of dewetting under the water-solvent mix; from isolated holes to the branched network to the droplets: (a–f) 18 nm thick PS film after 5, 12, 20, 30, 40 and 60 seconds; (g–l) 60 nm thick PS film after 5, 15, 90, 150 seconds, 5 and 30 minutes. (Scale bar: 20 μm)

Relatively thin films (< 20nm) show formation of long strands or branches from the unstable rims (Figure 1b) during the hole-growth (Figures 1c-1e). The instability of a hole-rim, which is derived from its cross-sectional curvature, has been observed and explained previously for dewetting in air.[12,13,15] A network structure of branched polygons is obtained by coalescence of holes (Figure 1d). The polygonal sides and strands eventually decay into droplets (Figures 1e, 1f). However, relatively thicker films (Figures 1g–l) do not show branching of polygons, the sides of which decay into droplets. The formation of slender branches or fingers in thinner films is indicative of higher cross-sectional curvature of rims,[12,13,15] as well as slippage[49] which has a stronger effect in thinner films.[49,50] It is known that fingering instability in a slipping straight edge produces droplets by a time-periodic disintegration of fingers.[49] However, disintegration of fingers was not observed *during the hole growth* in our experiments, but occurred largely after the formation of polygons. This is possibly because of a larger time scale for disintegration of fingers compared to the time scale for the hole growth and coalescence. The later decreases with an increased density of holes under the water-solvent mixture.

After dewetting was complete, the contact lines of droplets continued to recede for about an hour; after which there was no significant change. Even after 24 hours of immersing in the water-solvent mixture, no change in the number density of droplets was observed. Dewetted structures were examined in the FESEM in the transverse position to see the shape of the polymer droplet and its contact angle after various times of exposure to the water-solvent mixture. Figure 2 shows the time evolution of droplets of 1–2 μm in diameter that are generated from the dewetting of a 22 nm PS film.

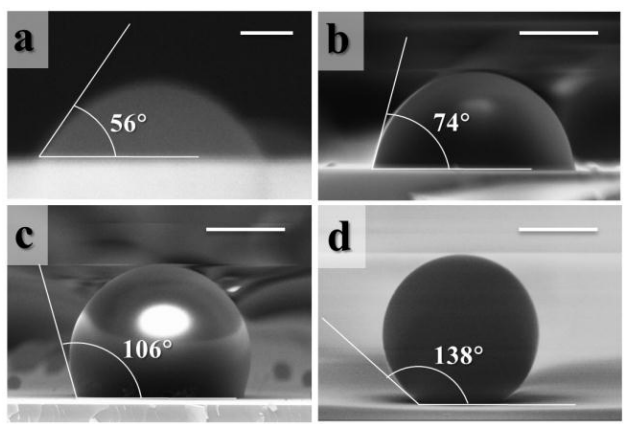

**Figure 2.** Time evolution of contact angles of dewetted droplets of 22 nm PS film showing large contact angles in case of dewetting in water-solvent mixture; (a) after 3 min (b) after 10 min, (c) after 20 min, (d) after 1 hour. (Scale bar: 500 nm)

The contact angle changes from 56° in 3 minutes to 138° after an hour. This is a tremendous increase in the contact angle from the dewetting in air which produces droplets with contact angles ~10°. As shown in figure 2 and discussed previously,[48] this method thus allows the fabrication of nano-lenses of controlled curvature in a wide-range. The slow increase in the droplet contact angle is of kinetic origin owing to the high viscosity of the polymer used above. This hypothesis was confirmed by working with a smaller molecular weight PS ($M_w$=2.5 kg mol$^{-1}$) of much smaller viscosity where the droplets attained the equilibrium configuration much more rapidly (~1 minute).

In order to probe the mechanisms of reduction in the droplet size, dewetting of PS thin films with thicknesses ranging from 7 nm to 92 nm was carried out in water-solvent mixture as well as in air saturated with the solvent vapors for quantitative comparison. The morphological evolution of instability was similar both by exposure to the solvent vapor in air and under the water-solvent mixture. Dewetting process, as described earlier, starts with the formation of isolated holes, which grow in numbers at first and then reach a maximum. The maximum number density of holes ($N_H$) is governed by the wavelength ($\lambda$) of long wave instability. In case of dewetting engendered by van der Waals attraction, $N_H$ and $\lambda$ theoretically scale with film thickness ($h$) as:[12–25] and $\lambda \sim h^{-2}$ and $N_H \sim \lambda^{-2} \sim h^{-4}$. Figure 3 compares these scaling obtained for dewetting in air and in the water-solvent mixture employed. Our data for dewetting in air by exposure to the solvent vapors are similar to thermally induced dewetting in air which has been extensively investigated earlier.[12,13,15,19,20] As shown in Figure 3a, dewetting in water-solvent mixture produces much higher hole density together with a relatively weaker dependence on the film thickness. For films thinner than 12 nm, full dewetting to form droplets in water-solvent mixture was observed almost instantaneously (<1 second) and hence number density of holes could not be precisely estimated. The least square linear fit on a double logarithmic plot gives $N_H \sim h^{-4.41\pm0.16}$ for dewetting in air and $N_H \sim h^{-3.02\pm0.12}$ for dewetting in the water-solvent mixture. Similarly scaling of $\lambda$ with the film thickness was found to be $\lambda \sim h^{-2.19\pm0.07}$ for dewetting in air and $\lambda \sim h^{-1.51\pm0.06}$ for dewetting under the water-solvent mixture (Figure 3b).

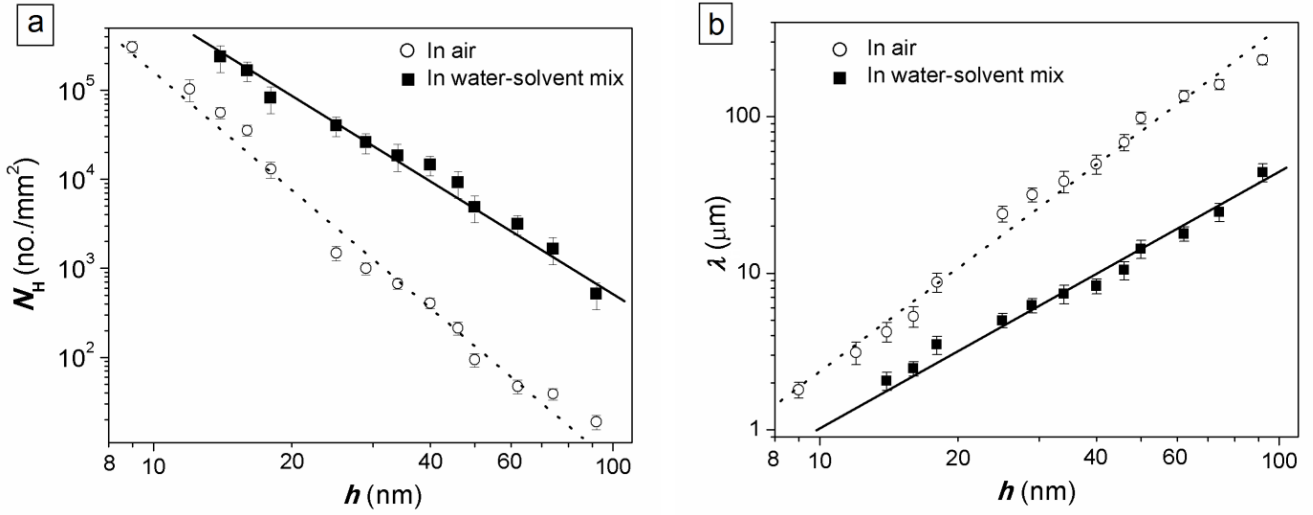

**Figure 3.** (a) Double logarithmic plot of the maximum number density of holes $N_H$ as a function of PS film thickness $h$; for dewetting in air and in water-solvent mix. The slope of the best fit lines are $-3.02\pm0.12$ in case of water-solvent mix and $-4.41\pm0.16$ in case of air. (b) Log-log plot of wavelength or the mean separation between holes as a function of PS film thickness for dewetting in air and in water-solvent mix. The slope of the best fit lines are $1.51\pm0.06$ in case of water-solvent mix as opposed to $2.19\pm0.07$ for air.

Figure 3 shows a regular dependence of instability wavelength on the film thickness in accord with the spinodal mechanism, equation (1), both for dewetting under air and under the water-solvent mixture, thus indicating that the spinodal mechanism can explain the observed dewetting on the flat substrates. The wavelength of instability relates to film thickness ($h$) and interfacial tension ($\gamma$) by equation (1), which in the case of van der Waals attraction, $\phi = A_e/6\pi h^3$ gives the following form: [12–15]

$$\lambda_{vw} = 4\pi\sqrt{\frac{\pi\gamma}{A_e}}h^2 \qquad (2)$$

Where $A_e$ is effective Hamaker constant. For dewetting of films in air, the exponent is indeed close to 2, as is also seen in several previous studies.[12–25] However, a weaker thickness dependence in case of dewetting in water-solvent mixture hints at a change in the destabilizing potential and not merely a reduction in the surface tension. Moreover, under water-organic mixture, the effective Hamaker constant is expected to be much smaller as compared to air, which should lead to bigger feature sizes and slower kinetics of dewetting. On the contrary, one observes an intensification of the instability. The data shown in Figure 3 clearly suggests a non-van der Waals potential of the form, $\phi \sim h^{-2}$, which together with equation (1) produces the exponents in accord with the data.

Equations (1) and (2) are valid for non-slipping or very weakly (slip length < film thickness) slipping films. Can strong-slippage account for intensification of instability and a reduced exponent for the wavelength? The effect of slippage on the *initial growth of instability and its wavelength* have been discussed in by Kargupta et al.[50] However, our data on $\lambda$ vs. $h$ for the dewetting under water-solvent mixture do not at all fit the theory of wavelength in slipping films. For a reasonable range of the effective Hamaker constant, 1 to $5\times10^{-21}$ J for the system (SiO$_x$/PS/Water-MEK), neither exponents nor numerical values could be fitted with any positive value of the slip length. In fact, slippage increases the wavelength of instability,[50] which together with a weakened van der Waals force under a liquid just cannot explain a decreased wavelength compared to air even when reduction in the surface tension is accounted for. Therefore, slippage is ruled out as the mechanism to alter the initial wavelength of instability and its dependence on the film thickness with an exponent of ~ 1.5.

Intensification of the destabilizing force field for dewetting of liquid PDMS films under water has also been observed previously.[51,52] This was attributed to the presence of electrostatic attraction,[51] for which, $\phi = \varepsilon\varepsilon_o U^2/2h^2$, which together with equation (1) predicts a different scaling:[51] $\lambda \sim h^{1.5}$ and $N_H \sim h^{-3}$. These predictions for the exponents are in agreement with our data (Figure 3) for dewetting under the water-solvent mixture, which indicates a possible electrostatic origin for the instability. In this case, the wavelength of instability varies with the film thickness ($h$) and interfacial tension ($\gamma$) by the following equation:[51]

$$\lambda_{el} = \frac{2\pi}{U}\sqrt{\frac{2\gamma}{\varepsilon_o}}h^{3/2} \qquad (3)$$

Where $U$ is the electrostatic potential difference (difference of work functions) across the film owing to different work functions of the three different media and $\varepsilon$ is the dielectric constant of the polymer thin film. In a recent article, Heier et al.[53] also considerd the role of unscreened electrostatic interactions in the instabilities of liquid-liquid bilayers of supramolecular assemblies where a reduced scaling exponent for the wavelength was observed.[53]

The value of interfacial tension of PS against the water-solvent mix is required for quantification of instability via equation (3). The measurement of interfacial tension was carried out using the pendant drop method. Figure 4 shows equilibrium pendant drop of PS in the solvent vapor-rich air and the water-solvent mixture. Experimental procedure and calculation of interfacial tension is provided as the supporting information. Surface tension of PS in solvent vapor rich air was found to be 25.8061±0.0235 mN m$^{-1}$, which is very close to the values reported earlier for the PS melt.[54,55] Whereas, the interfacial tension of PS and water-solvent mixture is obtained as 0.5507±0.0004 mN m$^{-1}$. Thus, there is about 50 fold reduction in the interfacial tension of PS in water-solvent mixture as compared to air.

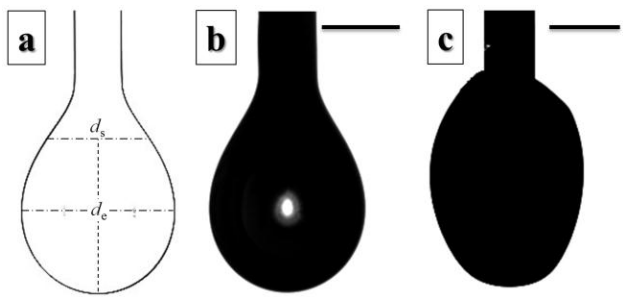

**Figure 4.** Measurement of interfacial tension by the pendant drop method. (a) Schematic diagram of pendant drop showing equatorial diameter ($d_e$) and diameter of a selected plane ($d_s$) at distance $d_e$ from the apex. (b) Pendant drop of PS in solvent vapor rich air. (c) Pendant drop of PS in water-solvent mixture. (Scale bar: 1 mm)

In case of dewetting in air engendered by van der Waals force, the value of $A_e$ was calculated from $\lambda_{vw}$ vs. $h$ data and found to be $1.76 \times 10^{-20} \pm 0.68 \times 10^{-20}$ J, which is in agreement with the previously reported values ($1.8 - 2.6 \times 10^{-20}$ J).[11,15,20]

Equation 3 can now be employed to calculate the best fit value of electrostatic potential difference causing the instability, which yields $U = 36$ mV. This magnitude of the potential appears to be realistic as in most polymeric-aqueous colloids and in oil-water emulsions, $U$ of the order of 10-100 mV.[51] The source of surface charge may be the unequal presence of dissociated $H^+/OH^-$ ions on the surface from water, which is the majority phase in the dewetting solution. However, more work is needed to precisely quantify the nature of electrostatic potential difference in a polymer-water/solvent mixture derived from the unequal adsorption of ions and/or difference of the work functions as originally suggested.[51] It is however certain that the explanation of the instability wavelength under the water-solvent mixture requires a non-van der Waals potential of the form, $\phi \sim h^{-2}$, which is also much stronger in strength than the effective van der Waals interaction of a solvent saturated polymer film under a water-solvent mixture.

Figure 5a compares mean droplet diameters for PS thin-films for dewetting in air and in the water-solvent mixture. It is evident that water-solvent mixture engenders droplets that are about one order smaller across the entire thickness range probed here.[48] Submicrometer size droplets were obtained upon dewetting of PS films thinner than 30 nm, which was not possible in dewetting in air for films even as thin as 10 nm. In addition, dewetting in the water-solvent mix engenders a weaker film-thickness dependence of the mean droplet diameter. This dependence, obtained by the linear fit on double logarithmic plot, was found to be $d_D \sim h^{1.26 \pm 0.07}$ for water-solvent mixture as opposed to the scaling, $d_D \sim h^{1.47 \pm 0.03}$ obtained for dewetting in air. The result for dewetting in air is in accord with the previous experiments[12,13] and theoretical expectation:[15] $d_D \sim h^{1.5}$ when dewetting is driven by the van der Waals force.[15] The weaker film-thickness dependence in case of water-solvent mixture can again be attributed to the weaker dependence of the electrostatic force on the film thickness as compared to the van der Waals force. Figure 5b, shows the number density of droplets ($N_D$) and the mean separation between them ($\lambda_D$) as a function of film thickness ($h$) for dewetting in the water-solvent mixture. $N_D$ and $\lambda_D$ show the similar dependence ($N_D \sim h^{-3.17 \pm 0.15}$ and $\lambda_D \sim h^{1.58 \pm 0.07}$) as in the case of holes for the thickness range 7–30 nm. However, for the films thicker than 30 nm, the thickness dependence weakens ($N_D \sim h^{-1.71 \pm 0.24}$ and $\lambda_D \sim h^{0.86 \pm 0.13}$), which possibly hints at the onset of a new-regime of dewetting for relatively thicker films which are known to be sensitive to the precise initial conditions and physico-chemical heterogeneities.[17,21,22,27,56,57] As is also shown in Figure 1, relatively thinner films form a more dendritic structure compared to the well-defined polygons in thicker films, which indicates a change in the process of droplet formation.

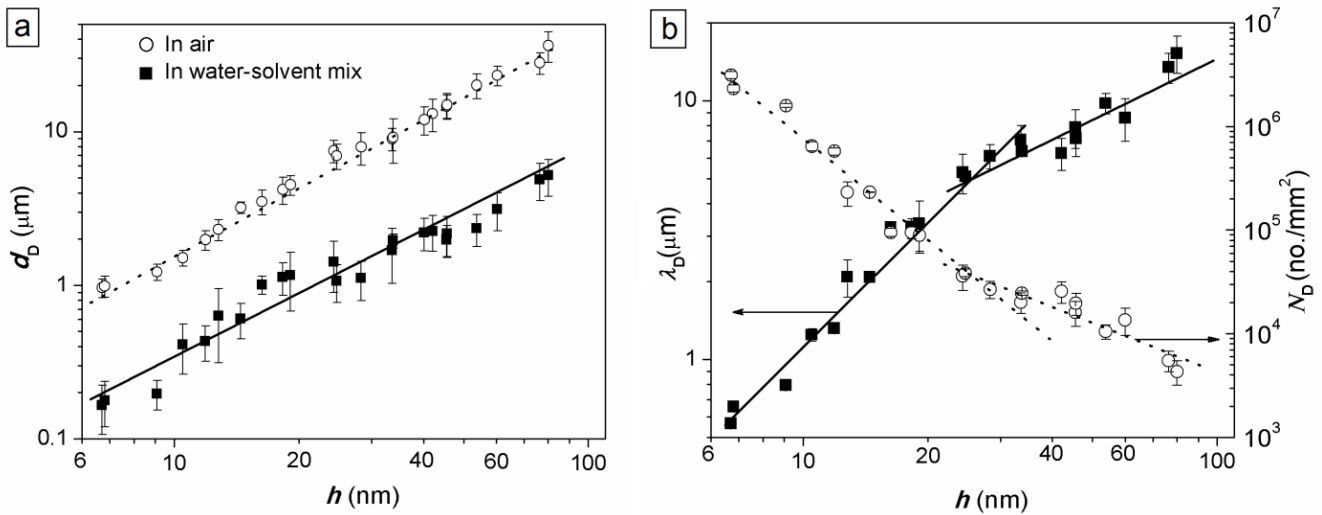

**Figure 5.** (a) Log-log plot of droplet diameter as a function of PS film thickness for dewetting in air and in water-solvent mix. The slope of the best fit lines are 1.26±0.07 in case of water-solvent mix as opposed to 1.47±0.03 in case of air. (b) Log-log plot of number density of droplets and the wavelength of fully dewetted structure as a function of film thickness in case of water-solvent dewetting.

**Dewetting on patterned substrates**

To demonstrate the potential use of dewetting under water-solvent mixture in self-organized patterning, dewetting experiments on the topographically micro-patterned substrates were carried out. The results are contrasted to dewetting in air to emphasize the power and usefulness of this technique for sub-micrometer patterning. Another important aspect considered here is that spin coating on topographically patterned substrates produces an uneven initial distribution of polymer thickness on the protruded and recessed parts. This initial non-homogeneity plays an important role in the understanding of the subsequent pattern development in dewetting. This aspect has not been considered in the earlier studies of dewetting on heterogeneous substrates.

1-D patterns in the form of open channels and 2-D patterns of array of pillars with circular or square cross-sections were fabricated on silicon by E-beam lithography (EBL). Size and pitch of these structures were varied in the range of 100 nm to 6 μm. PS thin-films were directly spin coated on patterned substrates, which results in slight thickness variation around the protruded features.[58–60] Spin coating on topographically patterned substrates largely produced uniform coating on the flat, bottom parts of the substrates, whereas the top portions of the channel walls and pillars received much thinner coatings that were even largely absent in the shorter channels and pillars of widths less than about one micrometer. For example, the AFM images of bare EBL patterned substrate in figure 6a shows the height of pillars ~115 nm. On coating with a PS thin-film under the conditions that would have produced a ~22 nm film on a flat substrate, a ~ 20 nm film was indeed coated on the flat regions of the substrate between the pillars. This was evidenced by the new height, ~95 nm, of the protruding pillars above the surface of the film with little or no coating on the pillars (Figure 6b).

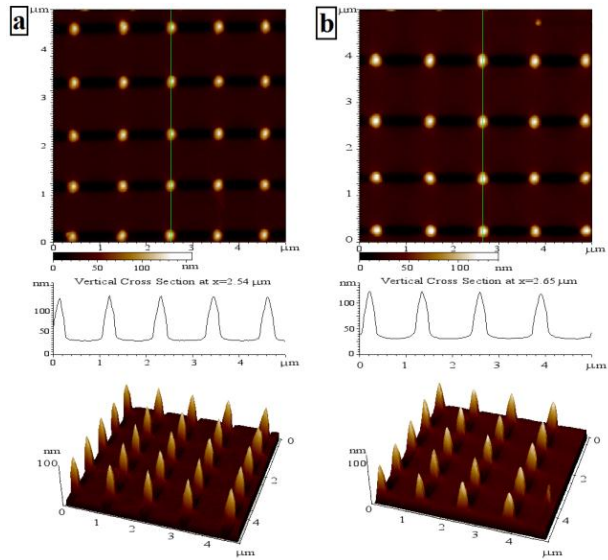

**Figure 6.** AFM images of EBL patterned substrates before and after coating thin PS film. (a) Bare EBL patterned silicon substrates; height of individual pillar is 115±5 nm, (b) ~20 nm thick PS coated EBL patterned silicon substrates, height of pillars above the base PS film is 95±5 nm.

In what follows, we first show that dewetting of thin films in air to produce droplets can be arrested in highly confined spaces owing to relatively large length scale of the instability. However, the same film can readily dewet to form droplet structures under the water-solvent mix. Figure 7 compares the structures produced on 1-D patterns (Figure 7c) by dewetting in air and under the water-solvent mixture. Images in the first column correspond to dewetting in air and the second column corresponds to dewetting in the water-solvent mixture. Third column images are the enlarged view of the second column images except in Figure 7c. Three different types of substrates were used namely: s1 (channel width, 1.7 μm; pitch, 2 μm); s2 (channel width, 150 nm; pitch, 400 nm) and s3 (channel width, 900 nm; pitch, 2 μm).

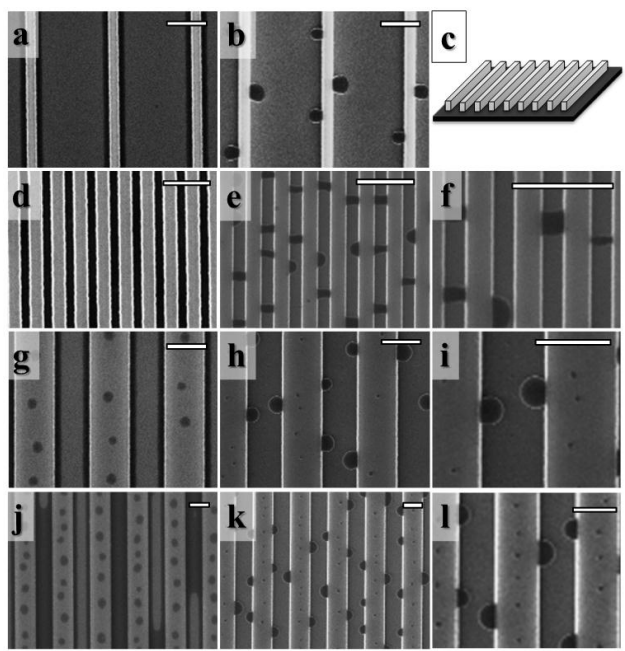

**Figure 7.** Dewetting of thin films in air and under water-solvent mix on substrates with microchannels. Images in the first column correspond to dewetting in air, while in the second and third columns are corresponding images for dewetting under water-solvent mix. (a–b) Dewetting of 9 nm thick PS film on a substrate with channel width of 1.7 μm and pitch of 2 μm, (c) schematic diagram of the topographically patterned substrate with parallel strips. (d–f) Dewetting of 9 nm thick PS film on a channel pattern with channel width of 150 nm and pitch of 400 nm. (g–i) Dewetting of 12 nm thick PS film on a channel pattern with channel width of 900 nm and pitch of 2 μm. (j–l) Dewetting of 14 nm thick PS film on a channel pattern with channel width of 900 nm and pitch of 2 μm. (Scale bar: 1 μm)

Figure 7a and 7b contrast the dewetting patterns produced by dewetting of a 9 nm thick film on the substrate s1. In case of dewetting in air, the polymer film in the channel breaks near its centerline and the entire polymer mass is eventually accumulated along the channel walls as a linear thread. However, when the dewetting of identical film was carried out in the water-solvent mixture, the polymer accumulated along the walls is in the form of droplets. When the same film was allowed to dewet in air on more closely spaced strips of substrate s2, polymer coating in the channels remained stable owing to its confinement in a space smaller than the length scale of instability (Figure 7d). However, in the water-solvent mixture, the film readily dewetted to produce droplets as seen in Figure 7e and its enlarged view in Figure 7f. Further, dewetting of a slightly thicker film (12 nm) on a 1-D pattern with larger channel-wall widths (substrate s3) produced droplets on top of the channel walls, both in air (figure 7g) and under water (figure 7h), in addition to the droplets in the channels which formed only under the water-solvent mix. However, the droplets formed under the mix were always much smaller than those in air. In air, dewetted polymer accumulated as a continuous thread along the walls (figure 7g). Figure 7j and 7k are the dewetted structures of a slightly thicker 14 nm thick PS film showing similar behavior as the previous case except that dewetting within the channels was largely suppressed in air, but proceeded under the water-solvent mix.

Dewetting in highly confined spaces is thus greatly facilitated by the use of water-solvent dewetting which allows for sub-micron patterning. However, 1-D substrates can only produce a linear ordering of dewetted droplets. A more highly ordered arrangement of droplets should be possible in 2-D confinement. Thus, dewetting of PS thin-films was investigated on EBL fabricated 2-D patterns of cylindrical and square pillars. Three different types of substrates were used namely: p1 (pillar diameter, 240 nm; pitch, 1.1 µm); p2 (pillar diameter, 540 nm; pitch, 2.4 µm) and p3 (pillar diameter, 1.1 µm; pitch, 5.7 µm). Figure 8, summarizes the results by highlighting the differences in air (first column) and water-solvent (second and third column) dewetting as a function of different levels of confinement imposed by the underlying 2-D patterns (Figure 8c). Figure 8a, 8b show the dewetted patterns of 12 nm PS film on the substrate p1. Dewetting in air shows rectangular patches of polymer formed by the rupture and pinning of the retreating polymer mass (Figure 8a). Dewetting of an identical film in the water-solvent mixture produces individual droplets that are associated with each pillar (Figure 8b). Qualitatively similar results were obtained for the dewetting of slightly thicker film of 14 nm on substrate p2 (Figure 8d–f). Greater mass of the film between the pillars in this case is able to produce more droplets, some of which are dispersed between the spaces between the pillars. For a 24 nm thick film dewetted under the mix, perfect ordering of the polymer droplets was obtained on the substrates p1 and p3 (Figure 8g–l). Interestingly, dewetting is arrested in this case on the substrate p1 when exposed to solvent vapors in air (Figure 8g), which is owing to a high confinement imposed by the substrate p1 ($\lambda_D^{air}$~24 µm, $d_D^{air}$~7.5 µm on flat substrate). The same film under water-solvent mixture produces a perfectly ordered structure where a droplet of about 410 nm is placed in-between every four pillars thus producing a square array of droplets separated by 1.1 µm (Figure 8h). This is a further 5 fold decrease in the droplet spacing along with more than a three-fold decrease in the droplet size as compared to the corresponding lengths on a flat substrate under water-solvent mix where $\lambda_D^{water}$~5 µm and $d_D^{water}$~1.4 µm. Thus, an important effect of substrate-topography is to further impose its lengthscale on dewetting when the substrate-pitch and instability wavelength become somewhat comparable. This effect is also known previously in the context of dewetting on topographically rough surfaces in air.[46] Figure 7i, shows the transverse view of the array of droplets. Further on substrate p3, dewetting of the same film in air produces droplets connecting 2 or 3 adjacent pillars (Figure 8j). In contrast, dewetting in water-solvent mixture again produces ordered structure with nearly 850 nm size polymer droplets separated by about 3 µm, which is again smaller as compared to the dewetting of such films on a flat substrate (Figure 8k–l). Thus, it is evident that the 2-D confinement can also result in a further reduction of the length scales of dewetting compared to a flat substrate. It is interesting to contrast dewetting shown in the group of Figures 8b, 8e and 8f to that in Figures 8h, 8k and 8l. In the former, one or two droplets are seen clinging to the pillars, but others droplets are randomly distributed in the spaces between the pillars (magnified view in Figure 8f). However, in Figures 8h and 8k (magnified view in Figure 8l), droplets are more precisely positioned in-between the pillars. These observations can be understood by a competition between the spinodal and 'nucleative' dewetting engendered by the pillars.[27,29,46,61] For the relatively thin (~ 12 nm) film in Figure 8b and 8e, the spinodal dewetting length scale (Figure 3) is smaller than the distance between the pillars, and thus, both the mechanisms can initiate dewetting: nucleative dewetting on the pillars and also spinodal dewetting between the pillars. As the film thickness increases, nucleative dewetting becomes much faster than the spinodal dewetting[27,29,61] and thus rupture and dewetting initiated from the pillars is dominant. The spinodal length scale is also higher than the pillar spacing in this case, so that dewetting between the pillars are suppressed. Thus, a periodic dewetting initiated by the pillars produces aligned droplets at the center of interstitial spaces. This transition to ordered dewetting as the film thickness is increased is also observed in air[46] and could be explained by simulations recently.[46,61]

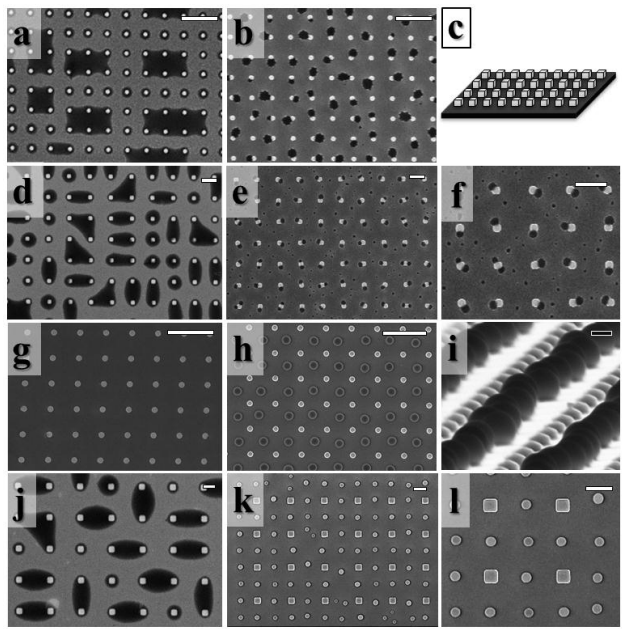

**Figure 8.** Contrast of the dewetting in air and in water-solvent mix on substrates with an array of pillars. Images in the first and second columns correspond to dewetting in air and dewetting under water-solvent mix, respectively. (a–b) Dewetting of a 12 nm thick PS film with dot size of 240 nm and pitch of 1.1 μm. (c) Schematic diagram of the topographically patterned substrate with cylindrical or square pillars. (d–f) Dewetting of 14 nm thick PS film with dot size of 540 nm and pitch of 2.4 μm. (g–i). Dewetting of a 24 nm thick PS film with dot size of 240 nm and pitch of 1.1 μm. (j–l) Dewetting of a 24 nm thick PS film with dot size of 1.1 μm and pitch of 5.7 μm. (Scale bar: 2 μm except (i) where it is 200 nm)

**Conclusions**

Field-induced self-organized patterning in ultra-thin polymer films produces structures that are limited to few microns to tens of microns in size because of the high energy penalty for the surface deformations on small scales and because of weak destabilizing van der Waals forces. We resolve here this long standing problem of miniaturization of length scales in thin film self-organization to sub-micron (~ 100 nm) levels by using a water-solvent mixture for dewetting. This proves to be a simple, powerful, flexible and inexpensive technique for the room temperature fabrication of sub-micron polymeric structures such as nano-lenses and their ordered arrays. We report more than an order of magnitude reduction in the feature size of the dewetted structures and thus fabricate structures as small as ~ 100 nm. The mean separation of these structures is also reduced close to 500 nm. Moreover, a far greater control over the shape and aspect ratio of the dewetted structures is demonstrated as the contact angles can be tuned in the range of 40-140 degrees. Controlled dewetting under the water-solvent mix on topographically patterned substrates is also employed for fabrication of sub-micron ordered structures. It is shown that under the same conditions, dewetting in air produces no dewetting, incomplete dewetting or is incapable of producing ordered structures owing to a large length scale of instability in air. A further miniaturization of length scales because of the 2-D confinement is also demonstrated.

We also explore the possible mechanisms of miniaturization of the interfacial instability and show that the thickness dependence of the characteristic length scales of dewetted structures is weaker ($\lambda \sim h^{1.51}$) as compared to the instability caused by van der Waals attraction ($\lambda \sim h^2$). Although a solvent saturated polymer film under the water-solvent mixture weakens the van der Waals destabilizing force, it greatly reduces the stabilizing interfacial tension and also apparently intensifies a destabilizing field of non-van der Waals origin with a potential of the form, $\phi \sim h^{-2}$, which could possibly be electrostatic.

In conclusion, dewetting under water-solvent mixture takes the physical self-assembly of polymer thin-films to its limits by producing sub-micron structures of various degrees of ordering and tunable shapes. Among other things, such patterns have the potential to be used as tunable polymeric micro-lenses and lens-arrays,[48] modifiers for the optical properties, and as delivery and positioning tools for a variety of functional materials that can be mixed with the polymer.

**Acknowledgements.** This work is supported by the Department of Science and Technology by an IRHPA grant and by DST Unit on Nanosciences. Authors acknowledge the initial efforts of Rabibrata Mukherjee and Partho Pattader in setting up some of the experiments.

**Supporting Information Available:** Experimental procedure for the calculation of interfacial tension of PS in solvent vapor rich air and water-solvent mixture. This material is available free of charge via the Internet at http://pubs.acs.org.